\documentclass[twocolumn,prb]{revtex4}
\usepackage{graphicx}
\usepackage{stmaryrd}
\usepackage{amsmath}
\usepackage{amssymb}
\usepackage{dcolumn}
\usepackage{bm}
\usepackage{color}
\begin{document}

%



\title{Functionalized Germanene as a Prototype of Large-Gap Two-Dimensional Topological Insulators}

\author{Chen Si$^1$, Junwei Liu$^1$, Yong Xu$^{1,2}$, Jian Wu$^1$, Bing-Lin Gu$^{1,2}$ and Wenhui Duan$^{1,2}$}
\email[\* Corresponding author. Email:\ ]{dwh@phys.tsinghua.edu.cn}
\address{$^1$Department of Physics and State Key Laboratory of Low-Dimensional Quantum Physics, Tsinghua University, Beijing 100084, People's Republic of China}
\address{$^2$Institue for Advanced Study, Tsinghua University,Beijing 100084, People's Republic of China}

\date{\today}

\begin{abstract}
We propose new two-dimensional (2D) topological insulators (TIs) in functionalized germanenes (Ge\emph{X}, \emph{X}$=$H, F, Cl, Br or I) using first-principles calculations. We find GeI is a 2D TI with a bulk gap of about 0.3 eV, while GeH, GeF, GeCl and GeBr can be transformed into TIs with sizeable gaps under achievable tensile strains. A unique mechanism is revealed to be responsible for large topologically-nontrivial gap obtained: owing to the functionalization, the $\sigma$ orbitals with stronger spin-orbit coupling (SOC) dominate the states around the Fermi level, instead of original $\pi$ orbitals with weaker SOC; thereinto, the coupling of the $p_{xy}$ orbitals of Ge and heavy halogens in forming the $\sigma$ orbitals also plays a key role in the further enlargement of the gaps in halogenated germanenes. Our results suggest a realistic possibility for the utilization of topological effects at room temperature.
\end{abstract}


\maketitle
\section{Introduction}
Recent years have witnessed many breakthroughs in the study of the topological insulators (TIs), a new class of materials with a bulk band gap and topologically protected boundary states \cite{Hasan,Qi}. Based on TIs, many intriguing phenomena, such as giant magneto-electric effects\cite{Qi2} and the appearance of Majorana fermions\cite{FuL}, are predicted, which would result in new device paradigms for spintronics and quantum computation. In particular, two-dimensional (2D) TIs have some unique advantages over three-dimensional (3D) TIs in some respects: all the scatterings of electrons are totally forbidden, leading to dissipationless charge or spin current carried by edge states; and the charge carriers can be easily controlled by gating.  Although many materials are theoretically predicted to be 2D TIs\cite{Kane1, ZhangSC, Murakami, Wada, Chao-xing, Xiao, Zhang, Fang}, so far only the HgTe/CdTe\cite{Konig} and InAs/GaSb\cite{Du} quantum wells are verified by transport experiments, which, however, still face particular challenges: very small bulk gap and incompatibility with conventional semiconductor devices. Therefore search and design of 2D TIs with larger gaps from the commonly used materials is indispensable for their practical utilization.

Graphene, with many superior properties from mechanical\cite{Lee,Si} to electronic\cite{Abergel, Si2}, has made remarkable progress in numerous applications. This has triggered extensive research on other 2D materials, such as silicene, germenene, tin monolayer, BN, MoS$_{2}$ and many others\cite{Cahangirov,Yaoyg,Tsai,Vogt,Xu,Song,Radisavljevic,Butler}. Among them, graphene and silicene could be well produced\cite{Novoselov,Vogt}, however, their practical applications as 2D TIs are substantially hindered by their extremely small bulk gaps ($10^{-3}$ meV for graphene\cite{Min} and 1.55 meV for silicene\cite{Yaoyg}). Germanene and tin monolayer have larger topologically nontrivial gap\cite{Yaoyg,Xu} but have not been fabricated experimentally yet. Very recently, germanane, a one-atom-thick sheet of hydrogenated germanene with the formula GeH, structurally similar to graphane, has been synthesized successfully\cite{Bianco}. With the predicted high mobility and easier integrability into the current electronics industry\cite{Bianco}, it is considered as a promising new star in the field of 2D nanomaterials\cite{Cui}. At the same time, the success of production of germanane has also stimulate the synthesis of its counterparts, such as halogenated germanenes.

In this work, using first-principles calculations, we investigate electronic and topological properties of single layer of hydrogenated/halogenated germanene (GeH, GeF, GeCl, GeBr and GeI). We find GeI is a 2D TI with an extraordinarily large bulk gap of about 0.3 eV, and GeH, GeF, GeCl and GeBr are trivial insulators but can be driven into nontrivial topological phases with sizable gaps larger than 0.1 eV under tensile strains.  We clearly reveal the physical mechanism for such large topologically nontrivial gaps: due to the functionalization of germanene, the $\sigma$ orbitals dominate the electronic states near the Fermi level, instead of original $\pi$ orbitals; and consequently the strong spin-orbit coupling (SOC) within $\sigma$ orbitals opens large nontrivial gaps.
Thereinto, the coupling of the $p_{xy}$ orbitals of Ge and heavy halogens in forming the $\sigma$ orbitals plays a key role in the further enlargement of the gaps in halogenated germanenes. The $\mathbb{Z}_{2}$ topological order is due to the $s$-$p$ band inversion at the $\Gamma$ point driven by the external strain or different chemical functionalizations.

\section{Models and methods}
Our calculations are performed in the framework of density functional theory with \emph{ab} \emph{initio} psudopotentials and plane wave formalism as implemented in the Vienna \emph{ab initio} simulation package \cite{VASP}. The Brillouin zone is integrated with a $18\times18\times1$ \emph{k} mesh. The plane-wave cut-off energy is set as
400 eV. The system is modeled by a single hydrogenated or halogenated germanene layer and a vacuum region more than 10 {\AA} thick to eliminate the spurious interaction between neighbouring slabs. The structures are relaxed until the remaining force acting on each atom is less than 0.01 eV/{\AA} within generalized gradient approximation (GGA) with
the Perdew-Burke-Ernzerhof (PBE) functional. Because GGA usually underestimates the band gap of germanide severely\cite{Broqvist}, we then use Heyd-Scuseria-Ernzerhof (HSE) screened Coulomb hybrid density functionals\cite{HSE} to calculate the electronic structures and $\mathbb{Z}_{2}$ topological invariant. The HSE calculations yield a band gap of 1.5 eV for the bulk GeH in a layered crystal structure, in good agreement with recent diffuse reflectance absorption spectroscopy measurement and theoretical calculation\cite{Bianco}.

\begin{figure}[tbp]
\includegraphics[width=0.3\textwidth]{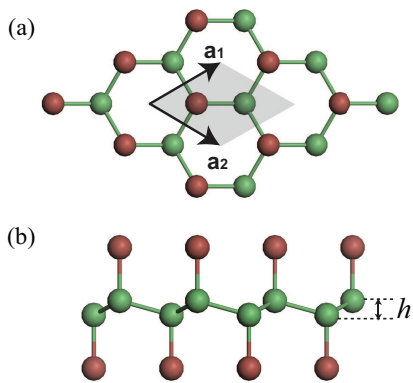}
\caption[]{(Color online) Top (a) and side (b) views of optimized structure of GeI displaying the primitive cell with Bravais lattice vectors a$_{1}$ and a$_{2}$ and the buckling of Ge plane $h$. Green and magenta balls represent Ge and I atoms, respectively. }
\end{figure}

\section{Results and discussion}
Figure 1 shows the optimized 2D GeI lattice structure, which is a fully iodinated germanene single layer. All the germanium (Ge) atoms are in \emph{sp}$^{3}$ hybridization forming a hexagonal network, and the iodine (I) atoms are bonded to the Ge atoms on both sides of the plane in an alternating manner. The equilibrium lattice constant is 4.32 {\AA}, with the buckling of the germanium plane ($h$), the Ge-Ge and Ge-I bond length being 0.69 {\AA}, 2.59 {\AA} and 2.57 {\AA}, respectively.

Without the SOC, GeI is gapless with the valence band maximum and the conduction band minimum degenerate at the Fermi level ($E_{F}$), as shown in Fig. 2(a). Including the SOC, a gap of 0.54 eV is opened at the $\Gamma$ point, along with an indirect gap of 0.3 eV (Fig. 2(b)).
\begin{figure}[tbp]
\includegraphics[width=0.5\textwidth]{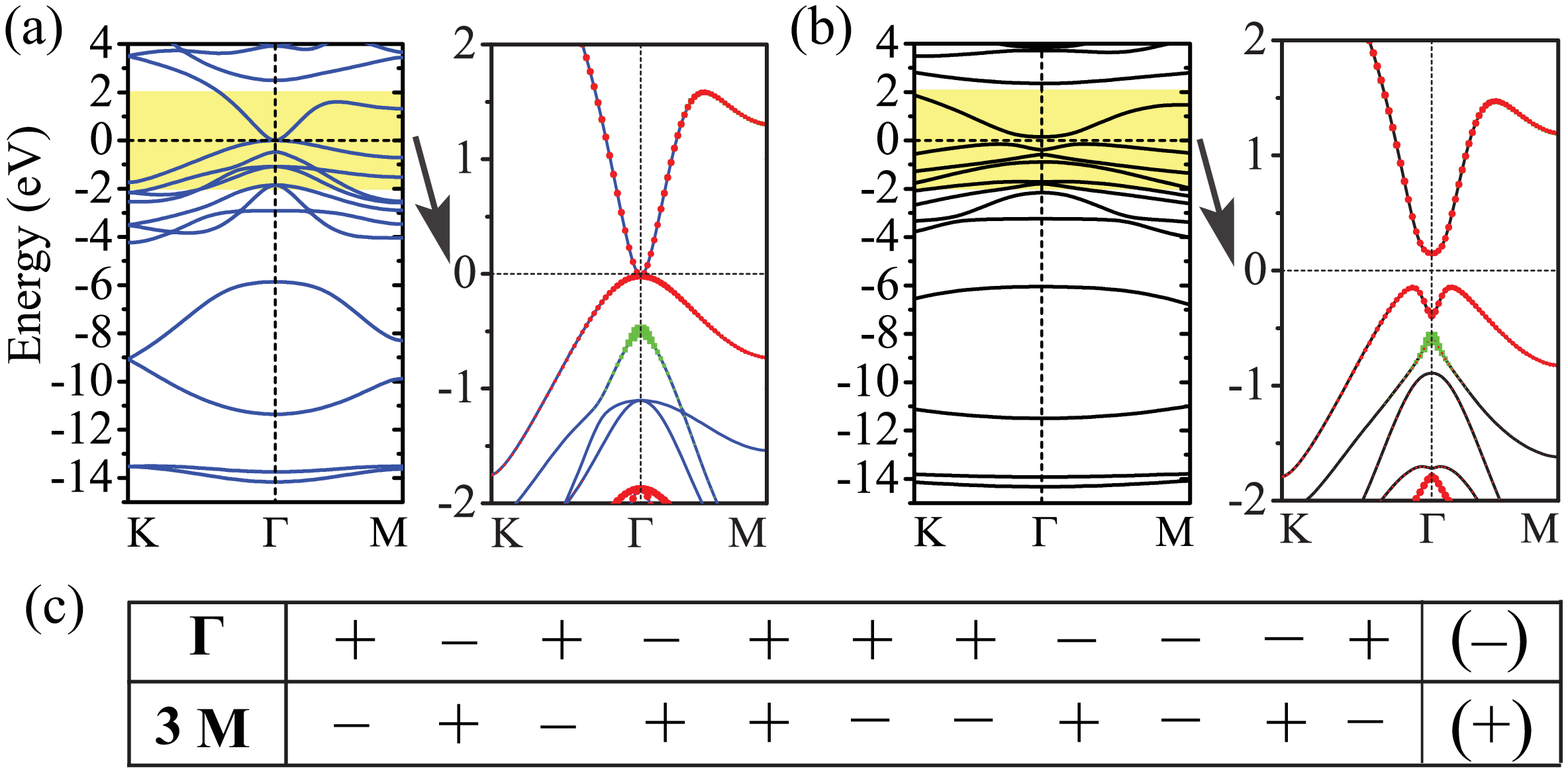}
\caption[]{(Color online) (a), (b) Band structures for GeI without SOC (blue line) and with SOC (black line) with zooming in the energy
dispersion near the Fermi level. The red circles and green squares represent the weights of the Ge-$s$ and Ge-$p_{xy}$ character, respectively. (c) The parities of eleven occupied bands at $\Gamma$ and three M points for GeI. The product of the parities at each $k$ point is given in brackets on the right.}
\end{figure}
To identify the 2D TI phase, a topological invariant $\nu$ is employed as ``order parameter'': $\nu=0$ characterizes a trivial phase, while $\nu=1$ means a nontrivial phase. Following the method proposed by Fu and Kane\cite{Fu2}, $\nu$ for GeI is calculated from the parities of wave function at all time-reversal-invariant momenta ($k_{i}$), one $\Gamma$ and three M points, as
\begin{equation}
\delta(k_{i})=\prod_{n=1}^{N}\xi_{2n}^{i}, \qquad (-1)^{\nu}= \prod_{i=1}^{4}\delta(k_{i})=\delta(\Gamma)\delta(M)^{3}, \nonumber
\end{equation}
where $\xi=\pm1$ denotes parity eigenvalues and $N$ is the number of the occupied bands. Fig. 2(c) shows the parities of eleven occupied bands at $\Gamma$ and M. It readily yields $\nu=1$, indicating quantum spin Hall effect can be realized in the single GeI layer.

For a 2D TI, a remarkable characteristic is an odd number of Dirac-like edge states connecting the conduction and valence bands. Thus we have also checked existence of the edge states in GeI. We use an armchair GeI nanoribbons with all the edge atoms passivated by hydrogen atoms to eliminate the dangling bonds. A large ribbon width of 9.3 nm is selected to avoid the interactions between the two edges. Fig. 3(a) shows the calculated electronic structure of GeI nanoribbon. One can clearly see the topological edge states (red lines) that form a single Dirac point at the $\Gamma$ point. Fig. 3(b) displays the real-space charge distribution of edge states at the $\Gamma$ point. It is visualized that these states are located at the two edges and distributed on not only Ge but also I atoms. The existence of edge states further indicates GeI to be a 2D TI. Moreover, its large bulk gap, about 0.3 eV, could be very useful for the applications of topological edge states in spintronic and computing technologies at room temperature.

\begin{figure}[tbp]
\includegraphics[width=0.5\textwidth]{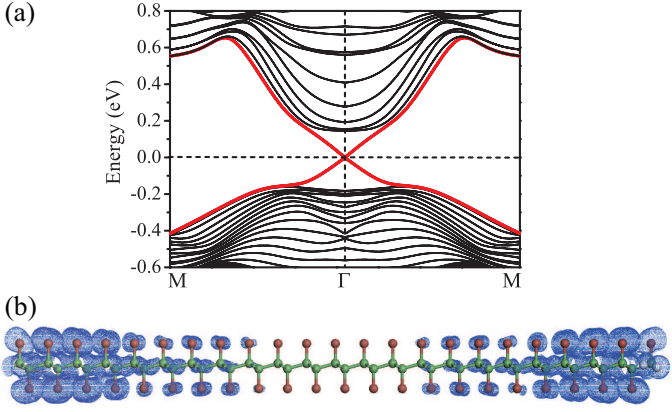}
\caption[]{(Color online) (a) Electronic structure for armchair GeI nanoribbons with the width of 9.3 nm. The helical edge states (red lines) can be clearly seen around the $\Gamma$ point dispersing in the bulk gap. (b) Real-space charge distribution of edge states at $\Gamma$.}
\end{figure}

The topological properties of GeI is closely related to the Ge-Ge bond strength, which is well confirmed by the direct comparison with GeH. GeH shares a similar geometric structure as GeI but has a smaller lattice constant (see Table 1). It is a normal insulator with a trivial gap of 1.60 eV (Fig. 4(a)), while tensile strain could drive it into a TI phase displaying a nontrivial gap of 0.20 eV at the $\Gamma$ point and an indirect bulk gap of 0.13 eV (Fig. 4(b)). Figs. 4(c) and (d) show the band evolution at the $\Gamma$ point of GeH under SOC and strain. Our calculations show that the states near $E_{F}$ are mainly contributed by the $s$ and $p_{xy}$ orbitals of Ge atoms, and thus we reasonably neglect other atomic orbitals in the following discussion. Firstly the chemical bonding of Ge-Ge makes the $s$ ($p_{xy}$) orbital split into the bonding and antibonding states, labeled with $s^{+}$ ($p_{xy}^{+}$) and $s^{-}$ ($p_{xy}^{-}$), where the superscripts $+$ and $-$ represent the parities of corresponding states. Without strain, $p_{xy}^{+}$ is lower than $s^{-}$, and the trivial gap ($E_{g}^{\Gamma}$) of the system (i.e., GeH) is just the distance between them (Fig. 4(c)). Applying tensile strain, with the Ge-Ge bonding strength weakened, the splitting of $s^{+}$ and $s^{-}$ ($\Delta s$) is rapidly reduced, causing $s^{-}$ shifting below $p_{xy}^{+}$ (Fig. 4(d)). In this inverted band structure, the $s^{-}$ is occupied, while the quadruply degenerate $p_{xy}^{+}$ is half occupied if the SOC is turned off, resulting in that $E_{F}$ stays at $p_{xy}^{+}$ level and the system becomes a semimetal. In contrast, turning on SOC, $p_{xy}^{+}$ is split into $\mid$$p, \pm3/2\rangle$ state with a total angular momentum $j = 3/2$ and $\mid$$p, \pm1/2\rangle$ state with a total angular momentum $j = 1/2$, thereby forming a nontrivial energy gap. It is also noted similar to this strain effects external pressure could induce band inversion and topological phase transition in some 3D systems\cite{PBarone}. From GeH to GeI, the different functionalization introduces the variation of electron density, which effectively induces a ``quantum electronic stress''\cite{Huhao}. Based on this concept, GeI should behave like a tensilely strained GeH, having the inverted band structure with $s^{-}$ lower than $p_{xy}^{+}$, as shown in Fig. 2(a) and (b).

\begin{figure}[tbp]
\includegraphics[width=0.5\textwidth]{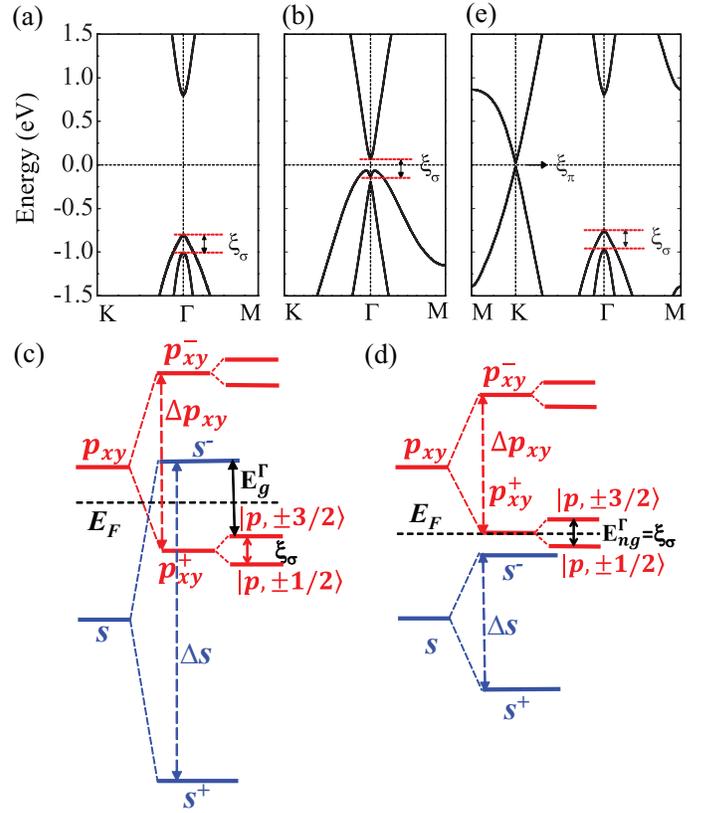}
\caption[]{(Color online) (a), (b), (e) Band structures with SOC for unstrained GeH, 12\% strained GeH and germanene, respectively. The splitting of $\sigma$ orbital at $\Gamma$ under SOC ($\xi_{\sigma}$), $\approx0.2$ eV. (c) and (d), schematic diagram of the evolution of energy levels at $\Gamma$ for GeH. Without strain, $p_{xy}^{+}$ is lower than $s^{-}$ (c). After applying enough large strain, $p_{xy}^{+}$ and $s^{-}$ are inverted (d). Under SOC, $p_{xy}^{+}$ is split into $\mid$$p, \pm3/2\rangle$ and $\mid$$p, \pm1/2\rangle$ states. }
\end{figure}

GeI, however, shows a much larger nontrivial gap at $\Gamma$ ($E^{\Gamma}_{ng}$) than strained GeH: 0.54 eV for the former, 0.20 eV for the latter. In GeH, according to a microscopic tight-binding model with similar basis and Hamiltonian as Ref.~\onlinecite{Castro}, we get the splitting of $\sigma$ orbital at $\Gamma$ under SOC ($\xi_{\sigma}$) which determines the size of $E^{\Gamma}_{ng}$ as shown in Fig. 4(c) and (d):
\begin{equation}
\xi_{\sigma}=(-3\delta+3\lambda+\sqrt{9\delta^{2}+6\delta\lambda+9\lambda^{2}})/4
\end{equation}
where $\delta=V_{pp\sigma}-V_{pp\pi}$, $V_{pp\sigma}$ and $V_{pp\pi}$ are hopping parameters corresponding to the $\sigma$ and $\pi$ bonds formed by 3$p$ orbitals; $\lambda$ is the SOC coefficient (H$_{\rm SO}$=$\lambda \overrightarrow{L}\cdot\overrightarrow{S}$)\cite{note}. It is noted that hydrogenation-induced corrugation has little effect on $\xi_{\sigma}$ and thus was ignored safely here. For GeH the $\sigma$ orbital at the $\Gamma$ point consists entirely of the Ge-$p_{xy}$ orbitals, so $\lambda$ could be approximately equal to Ge atomic SOC, $\xi_{\rm Ge}$.  Given that $\lambda/\delta$ is small enough ($\approx0.02$ for GeH), Eq. (1) is simplified based on the Taylor expansion:
\begin{equation}
\xi_{\sigma}=\lambda + \frac{\lambda^{2}}{3\delta}+o[\lambda/\delta]^{3}\approx \lambda\approx\xi_{\rm Ge}.
\end{equation}
From Eq. (2), we can see $\xi_{\sigma}$ in GeH is of the order of $\xi_{\rm Ge}$ (0.196 eV\cite{Yaoyg3}), in agreement with our DFT calculation (0.20 eV). Importantly, Eq. (2) also implies that $\xi_{\sigma}$ is almost independent of strain, thus the nontrivial gap in the TI phase is found to almost keep constant with the increase of strain. In GeI, the $\sigma$ orbital at the $\Gamma$ point is derived from the hybridization of the Ge-$p_{xy}$ and I-$p_{xy}$ orbitals. Here Eq. (1) still works but $\lambda$ in it should be the combination of Ge and I atomic SOC. Then the introduction of large I atomic SOC ascribed to its heavy atomic mass further increases dramatically the magnitude of $\xi_{\sigma}$ in GeI. Thus now we could easily understand why the nontrivial gap of GeI is much larger than that of GeH.

Actually, Eqs. (1) and (2) also work for germanene, silicene and graphene, where $\xi_{\sigma}$ is of the order of atomic SOC of Ge, Si or C. Fig. 4(e) shows the band structure of germanene. One can see its nontrivial gap is generated by the splitting of $\pi$ orbital at K under SOC ($\xi_{\pi}$), $\approx36$ meV, though the splitting of $\sigma$ orbital at $\Gamma$ is much larger, $\xi_{\sigma} \approx0.2$ eV. By comparing Fig. 4(a) with 4(e), it is clearly found that an important role of adsorbed atoms on germanene is to reduce the energy of $\pi$ orbital at the K point and induce the dominance of $\sigma$ orbital at the  $\Gamma$ point near the Fermi level. Thus we can use the larger SOC within $\sigma$ orbital to open a sizeable gap.

For many 2D materials like germanene, silicene, and graphene, the states around the Fermi level are generally contributed by $\pi$ orbitals. In order to obtain the TIs with visible topologically nontrivial gap, a conventional method is to increase the weak SOC of $\pi$ orbitals, such as applying compressive strain to increase the curvature of plane\cite{Huertas,Yaoyg}, regularly depositing heavy transition metals (TMs) on the surface to hybridize the $\pi$ orbital with the $d$ orbital of TMs\cite{Hu, Li}. Remarkably, our work provides a new alternative to increase the nontrivial gap, i.e., by making the orbitals (such as $\sigma$) with large effective SOC dominate the states around the Fermi level.

Recent theoretical work shows that SnI, iodinated tin monolayer, is also a 2D TI with a bulk gap of about 0.3 eV\cite{Xu}. However, the origin of this large gap of SnI was not clearly known. Similar to GeI, once the Sn mononlayer is iodinated, the original Sn $\pi$ orbital dominance near the Fermi level is changed into $\sigma$ orbital dominance, and then the larger SOC with $\sigma$ orbital introduces larger nontrivial gap. Given that Sn has a much larger atomic SOC than Ge, it is supposed that SnI would have a larger nontrivial gap than GeI. However, we observe that the bulk gap of GeI (0.3 eV) is unexpectedly comparable with that of SnI. We further find it is because the hybridization of Sn-$p_{xy}$ and I-$p_{xy}$ in forming $\sigma$ orbital is much smaller than that of Ge-$p_{xy}$ and I-$p_{xy}$. A simple orbital analysis indicates that the ratio of Sn-$p_{xy}$ to I-$p_{xy}$ component in the $\sigma$ orbital at the $\Gamma$ point is about $2:1$ while the ratio of Ge-$p_{xy}$ to I-$p_{xy}$ is about $2:3$.

\begin{table}[!tbp]
\caption{\label{tab:table1} The lattice constant($a$) and Ge-Ge bond length ($d_{\rm Ge-Ge}$) at equilibrium, critical strain ($\varepsilon_{c}$) and strained lattice constant($a_{c}$) where the topological phase transition occurs, nontrivial gap at $\Gamma$ ($E_{ng}^{\Gamma}$), indirect bulk gap ($\Delta$) for GeH, GeF, GeCl, GeBr and GeI.}
\begin{ruledtabular}
\begin{tabular}{cccccc}
system&GeH &GeF& GeCl &GeBr &GeI\\
\hline
$a$ (\AA) &4.09& 4.30 & 4.24 &4.25&4.32\\
$d_{\rm Ge-Ge}$ (\AA) & 2.47 & 2.55 & 2.54 & 2.55 &2.59 \\
$\varepsilon_{c}$ &10\% & 2\% & 3\% & 2\% & 0\%  \\
$a_{c}$ (\AA) &4.50 & 4.39 & 4.37 & 4.34 & 4.32  \\
$E_{ng}^{\Gamma}$ (eV) & 0.20 & 0.21 & 0.21 & 0.27 &0.54 \\
$\Delta$ (eV) & 0.13 & 0.13 & 0.13 & 0.18 &0.30 \\
\end{tabular}
\end{ruledtabular}
\end{table}

We further investigate other functionalized germanenes (GeF, GeCl and GeBr), structural analogues of GeI and GeH.  Similar to GeH, all of them undergo a phase transition from normal to topological insulators under tensile strain. Table 1 summarizes their lattice constants, Ge-Ge bond lengths, critical strains where phase transitions occur, and the nontrivial gaps in their TI phase. Due to the weaker Ge-Ge bond strengths in GeF, GeCl and GeBr, their critical strains are quite small, $\leq$ 3\%, indicating the experimental feasibility. Depending on different hybridization levels of the $p_{xy}$ orbitals of Ge and different halogens in forming $\sigma$ orbitals, they show different bulk gaps, however, all of which are larger than 0.1 eV, ample for practical application at room temperature. In addition, it is known that the nontrivial topologies of graphene, silicene and germanene are easily destroyed by the substrate, which breaks their \emph{AB} sublattice symmetry and introduces the trivial gap at the K point. In contrast, although all the topological properties of functionalized germanenes shown in this work are obtained for free-standing sheets,  their nontrivial topologies would be quite robust when they are on the substrate, because their band inversion occurs at the $\Gamma$ point rather than the K point and the full saturation of Ge $p_{z}$ orbitals ensures a weak interaction with the substrate.

\section{Conclusions}
In summary, based on first-principle calculations, we have studied the band topologies in functionalized germanene, including the recently synthesized germanene and halogenated germanenes. Among them, GeI is found to be a promising 2D TI with a very large gap of about 0.3 eV, while the others could be transformed into TIs with sizeable gaps larger than 0.1 eV by applying tensile strain. These large gaps are originated from strong SOC within the $\sigma$ orbitals, which is of the order of the Ge atomic SOC in GeH and further magnified in halogenated germanene due to the coupling between $p_{xy}$ orbitals of Ge and heavy halogens in forming $\sigma$ orbitals. The $s$-$p$ band inversion at the $\Gamma$ point, as the physical origin for the $\mathbb{Z}_{2}$ topological order, can be driven by different chemical functionalizations or the external strain. Our results clearly demonstrate the potential for utilization of topological edge states of germanium films in low-power spintronics devices at room temperature.

\section{Acknowledgments}
We acknowledge the support of the Ministry of Science and Technology of China (Grant Nos. 2011CB921901 and 2011CB606405), and the National Natural Science Foundation of China (Grant No. 11334006).


\end{document}